\documentclass[11pt]{article}
\usepackage[utf8]{inputenc}
\usepackage{hyperref}
\usepackage{graphicx}
\usepackage{amsmath,amssymb}
\usepackage[numbers,sort&compress]{natbib}

\title{Microgravity and Near-Absolute Zero: A New Frontier in Quantum Computing Hardware}
\author{Denis Saklakov \\
    Robotech Frontier Hub \\
    \texttt{ds@robotechfrontierhub.com} \\
    ORCID: \texttt{0009-0005-0756-7303}}
\date{}

\begin{document}
\maketitle

\begin{abstract}
Quantum computing qubits are notoriously fragile, requiring extreme isolation from environmental disturbances. This paper advances the hypothesis that a combination of microgravity and ultra-low temperature (near absolute zero) provides an almost “ideal” operating environment for quantum hardware. Under such conditions, gravitational perturbations, thermal noise, and vibrational disturbances are minimized, thereby significantly extending qubit coherence times and reducing error rates. We survey four leading qubit platforms – superconducting circuits, trapped ions, ultracold neutral atoms, and photonic qubits – and explain how each can benefit from a weightless, cryogenic setting. Recent experiments support this vision: Bose–Einstein condensates on the International Space Station (ISS) maintained matter-wave coherence far longer than on Earth, atomic clocks in orbit achieved record stability, and a photonic quantum computer deployed in space is demonstrating robust operation. Finally, we outline a proposed side-by-side experiment comparing identical quantum processors on the ground and in microgravity. Such a test would directly measure improvements in qubit coherence ($T_1$, $T_2$), gate fidelity, and readout accuracy when the influence of gravity is removed.
\end{abstract}


\section{Introduction}
Quantum computing hardware must operate under extraordinarily controlled conditions to preserve fragile quantum states. Even slight environmental disturbances -- thermal fluctuations, mechanical vibrations, electromagnetic noise, or gravitational forces -- can induce decoherence, causing qubits to lose information\cite{roomISS2023}. This article investigates the hypothesis that combining microgravity with ultracold (near-absolute-zero) temperatures creates an environment closer to “ideal” for quantum computation, one that mitigates not only gravity-related issues but a broad spectrum of noise and error sources. In essence, the extreme conditions of space (microgravity and cryogenic vacuum) are argued to improve qubit coherence times, suppress decoherence mechanisms, reduce gate and readout error rates, and enhance overall quantum computer performance beyond what is achievable in terrestrial labs.

We develop this argument through theoretical reasoning and examination of emerging experimental evidence. On the theoretical side, gravity's influence on quantum systems can be seen as a source of dephasing -- for example, spatially separated qubits accumulate relative phase via gravitational redshift, acting like a "noise channel"\cite{balatsky2025}. By eliminating weight and gravitational potential gradients, a microgravity environment may essentially remove this subtle decoherence channel. Moreover, operating at near $0$~K drastically suppresses thermal noise and blackbody radiation that would otherwise perturb qubits. Together, these conditions emulate an ideal isolated quantum system. We survey how four leading qubit platforms -- superconducting circuits, trapped ions, ultracold neutral atoms, and photonic qubits -- stand to benefit. We then review evidence from space-based quantum experiments (e.g., on the International Space Station), drop-tower microgravity tests, and temperature-dependent studies that support performance gains in these extreme environments. Finally, we propose an experimental design to directly test the hypothesis by comparing identical quantum processors in ground versus microgravity conditions, with careful isolation of gravitational effects from other factors.

\textit{Scope and Limitations:} This work is a conceptual survey and proposal, not a report of new experimental results. We synthesize theoretical insights and existing findings from the literature to argue for potential improvements in quantum hardware performance under microgravity and cryogenic conditions. The discussion is intended to be speculative yet grounded in known physics; any projected performance gains are based on extrapolation from cited experiments and models. Engineering challenges and uncertainties remain, and the hypothesis must ultimately be validated by dedicated experiments (such as the one proposed).

\section{Microgravity and Cryogenic Conditions as Qubit Environments}
Operating in microgravity (essentially free-fall) provides a uniquely quiescent environment where many perturbations present on Earth are minimized. In a weightless setting, there is no “up” or “down” pulling on equipment or particles; experimental apparatus can be more symmetrically designed and free of sagging or stress. For quantum devices, this means support structures and alignments remain more stable over time, potentially reducing slow drifts and vibrations that would miscalibrate qubits. For example, cold atom traps on Earth must constantly fight gravity to hold atoms, which requires steep trapping potentials and introduces asymmetry; in microgravity, the same atoms can float freely or be confined by much gentler, symmetric forces\cite{williams2024,gaaloul2022}. This eliminates one significant source of decoherence (the need for strong confining fields that can perturb quantum states) and allows atoms or ions to remain in place without levitation-induced noise. From a cryogenic perspective, near-absolute-zero temperatures freeze out almost all thermal excitations in the environment. Qubits at millikelvin temperatures are far less disturbed by thermal photons or phonons, and residual gas in the vacuum chamber condenses on cold surfaces, yielding an ultra-high vacuum with collision rates nearly zero\cite{wineland1998}. Combined, microgravity and an ultracold vacuum approximate an ideal isolated system: no convective air currents, negligible mechanical stress, minimal electromagnetic interference, and extremely low rates of gas collisions or blackbody-induced excitations. In such an environment, qubits can maintain superposition and entanglement significantly longer, improving their coherence lifetimes and the fidelity of operations.

\subsection{Gravity as Dephasing}
Beyond intuitive arguments, recent studies have formalized gravity's role as a decohering influence on quantum hardware. For instance, one analysis treated gravity as a pervasive dephasing field: in a static 1$g$ field, entangled qubits experience slight phase shifts due to gravitational potential differences, accumulating as a systematic decoherence over time\cite{balatsky2025}. Eliminating gravity (i.e., in free-fall) would remove this phase drift, unifying the proper time experienced by all qubits and thereby preserving relative phase coherence. In other words, microgravity can halt a “universal dephasing channel” that affects all quantum platforms\cite{balatsky2025}. Likewise, models of gravitational time dilation predict that a quantum superposition of an object at two heights will lose coherence as the internal clocks tick at different rates\cite{bassi2022}. Only by placing the entire system in the same gravitational frame (free-fall orbit or deep space) can one avoid such relativistic decoherence. These theoretical considerations strengthen the notion that gravity -- even static Earth gravity -- is more than just a trivial experimental inconvenience; it subtly but inevitably leaks information from quantum states into the gravitational field, unless we can remove or homogenize it. Microgravity thus offers a route to “turn off” one more source of qubit-environment entanglement.

\subsection{Cryogenic Vacuum \& Isolation}
Crucially, microgravity's benefits are intertwined with maintaining a cryogenic, vibration-free setting. Space is often thought of as cold and quiet, but it is not automatically so -- the cosmic microwave background is about 2.7~K (and even in low Earth orbit the effective thermal environment is $\sim$4~K)\cite{stackexchange2025}, and spacecraft are subject to solar heating and internal vibrational noise from pumps or fans. Reaching the $\sim10$~mK regime needed for superconducting qubits still requires active refrigeration even in space\cite{stackexchange2025}. Fortunately, space platforms can be equipped with closed-cycle cryocoolers or cryogen boil-off systems to achieve temperatures comparable to terrestrial dilution refrigerators (such technologies have been used on satellites for years, e.g. for space telescopes and particle detectors). What microgravity adds is a reduction in convective and gravity-driven effects in the cryostat: no buoyant hot spots or liquid stratification, and potentially easier alignment of multiple cooling stages without sag. In the weightless environment, cryogenic fluids and components can be arranged without concern for orientation, which may simplify some aspects of refrigeration engineering. More importantly, microgravity allows extreme vibration isolation techniques that are impossible on Earth. A spacecraft can be designed as a drag-free satellite -- using micro-thrusters to cancel out even minute forces -- such that an experimental package literally free-floats without touching the walls. This was demonstrated by missions like LISA Pathfinder, which achieved acceleration noise below $10^{-14}g$. In such a drag-free, 0$g$ environment, mechanical disturbances are vanishingly small, far below the seismic and acoustic noise floor of the quietest ground laboratory. For quantum computing hardware, which is highly sensitive to vibration, this offers an unprecedented stability of the reference frame.

In summary, microgravity and near-absolute-zero conditions together create a near-perfect “silence” across all decoherence channels: thermal, mechanical, and gravitational. In the following sections we examine in detail how specific qubit implementations stand to gain from these conditions, and what improvements have already been observed or predicted in practice.

\section{Superconducting Qubits in Microgravity}
\textbf{Current Challenges:} Superconducting qubits (such as transmon circuits) are typically operated at $\sim$10--20~mK in dilution refrigerators on Earth\cite{stackexchange2025}. At these temperatures, thermal excitations are suppressed and the superconducting state is stable, but qubit coherence is still limited by material defects, stray electromagnetic noise, and occasional high-energy radiation hits. Coherence times for state-of-the-art transmons have reached on the order of 0.1--0.3~ms as of 2025, but further improvements are stymied by “extrinsic” noise sources such as two-level system (TLS) defects in device dielectrics, fluctuating magnetic flux, and cosmic-ray-induced quasiparticles. Notably, vibrational and acoustic disturbances can couple into superconducting circuits through these defects or through microwave resonators (a phenomenon known as microphonics)\cite{ivezic2025}. For example, a 2024 study found that mechanical vibrations from a pulse-tube cryocooler shook a qubit chip enough to perturb ensembles of TLS defects, causing bursts of correlated qubit errors\cite{ivezic2025}. Even though an electron in a superconducting film is not directly sensitive to 1~$g$ of gravity, the indirect effects -- stress on the chip, moving parts in the fridge, or vibrations detuning cavity frequencies -- can degrade performance. Moreover, as superconducting qubits become extraordinarily coherent, they start to notice subtle effects like Earth's gravity: one proposal showed that entangled supercurrents in a transmon would accumulate a gravitational phase difference over time, providing a measurable qubit dephasing unless corrected\cite{balatsky2025}.

\textbf{Microgravity Benefits:} Placing a superconducting quantum processor in microgravity could alleviate several of these issues. First, the elimination of weight means the cryostat and qubit chip experience no sag or differential stress -- delicate microwave cavities and Josephson junctions will not deform under their own weight. This may reduce strain-induced TLS activation in materials. Vibration isolation is vastly easier when the entire experiment can free-float: a cryostat in orbit can be mounted on soft springs or even magnetic suspension without having to support its own mass, leading to superb isolation from any vehicle jitter. In principle, a drag-free satellite could allow the qubits to experience essentially no external acceleration or vibration at all. This matters because even tiny vibrations (picometer-scale) can modulate qubit frequencies via Stark or strain effects. Microgravity also removes the need for rigid mechanical support structures that conduct acoustic noise; the whole setup can be more compact and freely floating. As a result, microphonics that currently plague superconducting qubits would be greatly suppressed, yielding more stable qubit frequencies and longer $T_2$ coherence times. For instance, the noise from pulse-tube coolers could be better isolated or canceled out in microgravity, preventing the vibration-induced TLS fluctuations that cause qubit error bursts\cite{ivezic2025}.

Another advantage is uniform cryogenic cooling. In microgravity, cooling can rely purely on conduction and radiation without convective heat flow. This might enable more uniform ultra-low temperatures across a large qubit chip. On Earth, gravity can cause slight temperature gradients (warm air rising, cold liquid helium sinking) which create local hot spots or fluctuating temperatures. A qubit in orbit, thermally anchored to a radiator facing deep space (~3~K background), can maintain an ultra-stable base temperature. Thermal stability is crucial because residual thermal photons in readout resonators or control lines are a major decoherence source for superconducting qubits; even at 20~mK a resonator can harbor stray microwave photons unless properly thermalized. A colder, more stable environment means fewer of these stray photons and hence fewer spurious qubit excitations induced by them.

Microgravity also enables better mitigation of cosmic ray impacts. Paradoxically, space has a higher flux of high-energy radiation (cosmic rays, solar particles) than Earth's surface, and such radiation is known to break Cooper pairs and create quasiparticles that momentarily destroy a qubit's superconducting coherence. However, in a dedicated quantum satellite one can incorporate thick radiation shielding (e.g. lead or polyethylene layers) without concern for weight. A heavy shield that would be impractical on Earth (due to sag or structural support needs) can be included in a spacecraft to dramatically cut down the flux of ionizing particles reaching the qubits. Additionally, microgravity allows consideration of novel qubit architectures that exploit the environment -- for example, aligning qubits so that any cosmic ray that does penetrate hits multiple qubits identically (a common-mode hit that error-correcting codes could detect). Without gravity, even the orientation and configuration of the qubit array are more flexible for such optimizations.

Finally, by removing gravitational redshift and time-dilation differences across the device, all superconducting qubits on a chip in free-fall share the same inertial frame and proper time. This means a large-scale superconducting processor (which might span several centimeters) would no longer have tiny elevation-dependent frequency offsets. While this effect is minuscule (a 1~cm height difference on Earth yields a fractional frequency shift on the order of $10^{-18}$ due to gravitational potential), it is not entirely academic when qubits approach $10^{-3}$~s coherence times and $10^{-4}$ error rates -- such tiny biases can accumulate or interfere with error-cancellation schemes. A microgravity quantum computer would operate in a pure reference frame free of terrestrial gravitational perturbations\cite{balatsky2025}.

In summary, superconducting qubits in microgravity are expected to exhibit longer coherence times and more stable operation due to the removal of vibration-induced noise, the absence of static gravitational dephasing, and the potential for enhanced shielding and thermal stability. Classical gravitation has a nontrivial influence on quantum computing hardware (as one study noted\cite{balatsky2025}), so eliminating gravity can push that influence effectively to zero. Overall improvements might manifest as reduced dephasing (with $T_2$ approaching the intrinsic limits set by materials), lower stochastic error rates (especially those correlated with vibrations or cosmic events), and the ability to perform deeper quantum circuits before decoherence accumulates.

\section{Trapped-Ion Qubits in Microgravity}
\textbf{Current Challenges:} Trapped-ion quantum computers achieve some of the longest qubit coherence times today -- hyperfine-state ions can maintain coherence for minutes or more in well-shielded setups, and multi-qubit entangling gate fidelities above 99.9\% have been demonstrated. However, these impressive feats require extreme isolation: ultra-high vacuum (to prevent background gas collisions) and very stable electromagnetic trapping fields. On Earth, ions are held in radio-frequency (RF) Paul traps or static Penning traps where electric fields must counteract gravity. While a single trapped ion's weight is astronomically small, gravity can still affect ion traps in practical ways. For example, most linear RF traps are oriented horizontally so that gravity causes a constant slight displacement of the ion from the RF null; this is compensated by DC fields, but any fluctuation in that compensation (due to electrode vibration or voltage noise) will shake the ion. In vertical trap orientations, gravity will directly pull ions out of the trap potential if it is not sufficiently strong. Thus, trap parameters on Earth must be tuned to be “tight” enough to overcome gravity. This typically means higher trap frequencies (stronger confinement) than would otherwise be needed to simply contain the ion's thermal motion. Strong confinement is beneficial for stability but has downsides: the ion's motional ground-state wavefunction is very small and the heating rates can be higher due to electric field noise that scales with trap frequency. Moreover, laser-based gates require the ion's motion to be well controlled; excess micromotion or oscillation from trap imperfections (often exacerbated by gravity pulling the ion off-center) can reduce gate fidelity. Mechanical vibrations of the trap structure or optics are also problematic: even nanometer-scale vibrations modulate the distance between ions and laser beams or electrodes, causing phase errors in gate laser pulses\cite{ivezic2025}. Laboratories combat this by using pneumatic isolation tables and active stabilization, but some vibrational noise (e.g. building rumble or acoustic noise) is unavoidable on Earth. Background gas collisions, though infrequent in $10^{-11}$~bar vacuums, do occur and will instantly decohere an ion's qubit state via momentum transfer. At room temperature, blackbody radiation also induces spontaneous transitions or level shifts (Stark shifts), imposing a limit on the accuracy of optical clock qubits. Cryogenic ion traps (operating at 4~K or below) have shown dramatically reduced background gas and electric field noise, enabling ion storage for days\cite{wineland1998} and lower motional heating rates. However, most multi-ion quantum computers still run at ambient or only moderately low temperatures (tens of kelvin) due to engineering complexity.

\textbf{Microgravity Benefits:} A microgravity environment provides immediate relief from one of the ion trap's constraints: the need to support ions against weight. In orbit, an ion does not “fall,” so even an extremely weak trapping potential can confine it relative to the trap center (which co-moves in free-fall). This means one could operate traps at much lower frequencies without losing the ion -- an approach not possible on Earth, where too weak a trap would let the ion drop out. Weaker trap potentials are advantageous because they reduce driven micromotion and RF-induced heating. The ion can occupy a larger, more forgiving region of the potential well. With gravity out of the picture, the trapping fields can be made perfectly symmetric and centered, eliminating asymmetry that leads to micromotion. (As noted for neutral atoms as well, in microgravity the need for a relatively strong potential to support particles is removed\cite{williams2024}.) In a microgravity ion trap, an ion could literally be confined by the tiniest “nudges” of electric field -- just enough force to gently bounce it within a small region -- which minimizes the driven motion at the RF frequency. This in turn lowers the heating rate from electric-field noise (often observed to scale strongly with trap frequency). The outcome would be longer motional coherence and easier ground-state cooling, benefiting two-qubit gate fidelity (which depends on maintaining coherent motional states).

Microgravity also allows far more freedom in trap geometry and placement. Ions could be held in very large traps or even multiple trap modules separated by macroscopic distances, without concerns of sag or misalignment due to gravity. A network of ion traps on a space platform could float freely relative to each other, linked by optical fibers or free-space lasers, without needing heavy support structures that introduce vibration. Even within a single trap chip, microgravity eliminates the small force that can cause the chip to flex or the trap assembly to strain over time. This mechanical stability means the electric field environment of the ion is more stable, leading to fewer fluctuating stray fields that cause qubit frequency noise. Any residual mechanical vibrations present on the spacecraft (from pumps, etc.) can be isolated more effectively as well -- as discussed earlier, a free-flying experiment can achieve superb vibration isolation. In an ion trap, vibrations of optical tables or trap electrodes on Earth can directly modulate the trapping fields, causing phase and amplitude noise\cite{ivezic2025}. In microgravity, the whole trap and its optics could be mounted on a single rigid platform that drifts freely, so external vibrations impart nearly no relative motion between trap and laser beams. Notably, the ISS Cold Atom Lab demonstrated this principle: they measured atomic interferometer phase noise caused by vibrations, and by characterizing it (the ISS had some residual vibration) they could anticipate the needs for future quiet platforms\cite{swayne2024}. With advanced isolation, a space-based ion system could ensure that, for example, a 1~nm vibration of a mirror (which would be disastrous for a phase-stable laser gate) either does not occur or is common to both ion and laser (thus canceling out as common-mode).

Another major benefit of space for ion qubits is the availability of an extended high-quality vacuum. Space itself is an ultra-high vacuum environment, and while an ion trap apparatus must still have an enclosure to maintain cryogenics and keep out contamination, the baseline vacuum achievable is extremely high. Furthermore, microgravity eliminates convective flows that in Earth labs can stir residual gases and increase the likelihood of collisions. A cryogenic ion trap in space, with getters and cryopumps, could likely reach pressures below $10^{-15}$~bar. At that level, an ion might not collide with a gas molecule for years. Indeed, in a cryogenic trap on Earth, Hg$^+$ ions were stored for many days without loss\cite{wineland1998}. In microgravity one can expect similarly negligible collision rates -- which translates to essentially zero interruptions of ion coherence from background gas. Additionally, at near 0~K, the blackbody radiation field is negligible. For example, if one uses an optical $^{171}$Yb$^+$ clock qubit, blackbody radiation at 300~K causes Stark shifts and a finite lifetime of the metastable state; at 4~K these effects are almost gone, and at 0~K they vanish entirely. Thus qubit frequency stability and coherence improve. (Microgravity by itself does not change temperature, but the space environment makes it easier to maintain a steady cryogenic state through passive radiative cooling supplemented by active cooling.) The combination of space and cryogenics removes two significant decoherence factors for trapped ions: gas collisions and thermal radiation.

Microgravity offers practical operational benefits as well. For instance, in microgravity one could co-trap different ion species without differential settling. In quantum computing, dual-species traps (one ion species for logic, another for sympathetic cooling) are often used. On Earth, ions of different mass in the same trap will have slightly different equilibrium positions (heavier ions sit lower in an axial trap under gravity). In microgravity, an ion crystal has no preferred orientation, so mixed-species arrays remain perfectly overlapped, potentially simplifying multi-species quantum logic operations. This scenario was demonstrated in part by the CAL experiment on the ISS, which simultaneously manipulated two different atomic species in free-fall\cite{williams2024}.

Finally, trapped-ion technology has already proven itself in space in the form of atomic clocks. NASA's Deep Space Atomic Clock mission flew a mercury-ion trap clock in low Earth orbit in 2019. The Hg$^+$ trap operated successfully, demonstrating a fractional frequency stability of about $3\times10^{-15}$ at one day\cite{alonso2022}. Its success showed that microgravity did not introduce any new decoherence -- if anything, the clock's limiting noise was technical (electronics and reference instability) rather than gravitational. Similarly, a cold rubidium atomic clock on China's \textit{Tiangong-2} space lab (the CACES experiment) achieved $3\times10^{-13}/\sqrt{\text{s}}$ stability under microgravity, proving the robustness of laser-cooled trapped atom/ion techniques in orbit\cite{alonso2022}. These pathfinders indicate that a trapped-ion quantum computer could function in space -- and with microgravity, potentially surpass ground performance by operating at lower trap frequencies and experiencing fewer interruptions. In short, microgravity permits gentler trapping and a quieter environment for ion qubits, which should lengthen coherence, reduce motional decoherence during multi-qubit gates, and yield higher gate fidelity and stability.

\section{Ultracold Neutral Atom Qubits in Microgravity}
\textbf{Current Challenges:} Neutral atoms (including cold atom qubits in optical lattices or arrays of optical tweezers, and atoms used in Bose--Einstein condensates for quantum simulation) are exquisitely sensitive to gravity. Unlike ions, neutral atoms have no net charge to trap with static fields; they are confined by electromagnetic fields (laser light or magnetic gradients) that must compete directly with gravity's pull. In a typical cold atom experiment on Earth, if one simply turns off the trapping potential, the atoms will fall under gravity and leave the interaction region in a fraction of a second (often only a few tens of milliseconds for millimeter-scale setups). Even while trapped, gravity can distort the trapping potential -- for instance, in a harmonic magnetic trap, gravity shifts the equilibrium point downward, resulting in an anharmonic “sag” in the potential. This breaks the symmetry and causes density gradients in an atomic ensemble. Optical lattices (standing light waves) aligned horizontally will have atoms rattling against the lower potential wells due to gravity. The finite free-fall time on Earth fundamentally limits experiments that require long interrogation times or free evolution of matter waves; phenomena that take more than a few hundred milliseconds to unfold are difficult to observe because the atoms hit the container walls or leave the laser beams. Additionally, neutral atom qubits often rely on extremely low temperatures (nano- to picokelvin) to achieve long coherence. But reaching such temperatures via evaporative cooling takes time, during which gravity is tugging at the atoms. Many cold atom experiments become a race against the clock to cool and utilize the atoms before gravity spoils the trap. Mechanical vibration noise also plagues neutral atom setups: vibrations of mirrors or optical tables impart phase noise to optical traps and interferometers\cite{ivezic2025}. For example, if a mirror that reflects a trapping laser beam vibrates, the entire optical lattice shakes and atoms see a fluctuating potential, causing decoherence. Similarly, acoustic noise can modulate refractive indices or drive currents in magnetic coils, adding noise to atomic qubits. Another challenge is background gas collisions (though cold atoms are typically in $10^{-9}$--$10^{-10}$~bar vacuums, collisions still truncate condensate lifetimes to a few seconds). Blackbody radiation at room temperature can photo-ionize ultra-cold atoms or drive unwanted transitions (especially for highly sensitive Rydberg-state qubits, which are very susceptible to ambient radiation).

\textbf{Microgravity Benefits:} Microgravity is a game-changer for neutral atom quantum systems. The most direct benefit is the dramatic extension of free-fall time. In microgravity, an ultracold atomic cloud can float essentially indefinitely (limited only by vacuum quality and atomic interactions), allowing experiments to run orders of magnitude longer. On the International Space Station, NASA's Cold Atom Laboratory (CAL) created Bose--Einstein condensates (BECs) of $^{87}$Rb and was able to observe them freely expand for over a second, whereas on Earth the condensate would collide with the trap walls in a few tens of milliseconds\cite{roomISS2023}. In orbit, microgravity allows scientists to watch a condensate expand for seconds rather than fractions of one, as the freefall of the atoms is indefinitely long unlike on Earth where gravity causes the condensate to be shifted out of the trap region. That extended observation time directly translates to longer coherence times for matter-wave qubits. Experiments have demonstrated matter-wave interference in microgravity lasting up to an order of magnitude longer than on the ground\cite{williams2024,roomISS2023}. In one case, interference fringes were observed for over 150~ms in free expansion in CAL's compact chamber -- an unprecedented duration for atomic interference in such a setup.

Microgravity also allows extremely low effective temperatures to be achieved -- regimes very hard to reach on Earth. The CAL facility and other space experiments use techniques like delta-kick collimation (a method to reverse and slow atomic expansion) to reach effective temperatures of only tens of picokelvin\cite{gaaloul2022}. For example, CAL reported expansion energies below 100~pK by letting a BEC expand in microgravity and then applying gentle focusing kicks\cite{gaaloul2022}. These picokelvin ensembles have such low kinetic energy that atoms scarcely move relative to each other. On Earth, gravity would pull even these ultra-cold atoms out of the trap region, but in microgravity they remain available for manipulation. Why does this matter for quantum computing? In proposals for neutral atom qubits (like arrays of atoms in optical tweezers performing gate operations via Rydberg excitation), lower temperatures mean the atoms are more localized and less perturbed, which improves gate fidelity. Temperature broadening of atomic transition frequencies (Doppler and recoil effects) is minimized, so laser-driven qubit rotations can be more precise. If one can routinely prepare qubits at, say, 50~pK instead of 1~µK, the coherence of internal states and the stability of clock transitions is improved, and the chance of losing atoms from the trap during operations is essentially zero. Microgravity thus helps achieve the ultracold initial conditions that quantum logic with neutral atoms thrives on.

A weightless environment also means trapping potentials can be perfectly symmetric and subtle. Magnetic or optical traps no longer have to compensate for a constant downward force, which often introduces anharmonic terms. In microgravity, one can create undistorted, symmetric traps for atoms\cite{gaaloul2022}, enabling novel configurations like shell-shaped BECs or uniform-density optical lattices that would be impossible under 1~$g$. Indeed, scientists using CAL created a freely expanding “bubble” BEC -- a hollow spherical shell of atoms -- which can exist only without gravity (otherwise it would sag under its own weight)\cite{gaaloul2022}. Such symmetric, gentle trapping conditions are ideal for quantum simulation and potentially for quantum computing, because they reduce inhomogeneity across an array of qubits. Every atom can experience the same trapping frequency to high precision, simplifying multi-qubit gate tuning. Also, the absence of a preferred direction (no gravity) means one could arrange neutral atom qubits in truly three-dimensional architectures (not just a pancake or line on a horizontal surface). This could be useful for scaling up the number of qubits or implementing 3D error-correcting code layouts, which are hard to maintain on Earth due to gravity-induced density gradients.

Microgravity also suppresses one of the invisible decoherence sources: differential gravitational potential during quantum operations. A fascinating possibility is to perform, for example, an atomic clock comparison between two heights. On Earth, even a 30~cm height difference causes a measurable phase shift from gravitational time dilation (about $3\times10^{-17}$ fractional frequency offset)\cite{bassi2022}. In a quantum algorithm where an atomic qubit is delocalized between two positions (like in a large-scale atom interferometer), this effect could introduce a phase error. In microgravity or space, that effect either vanishes (in freefall) or can be tuned by orbital maneuvers. Essentially, all atoms share the same gravitational frame, which keeps their quantum phases synchronous.

Another huge benefit is the elimination of convective and acoustic disturbances around the atoms. In Earth labs, residual vibrations -- from HVAC systems, nearby traffic, even people walking -- can shake optics and cause phase noise on laser beams controlling neutral atoms\cite{ivezic2025}. In a properly designed spacecraft, with no footsteps or traffic and the experiment isolated, these noise sources can be orders of magnitude lower. Laser frequency stability can also improve in microgravity: ultra-stable reference cavities for lasers often suffer slight deformations due to gravity; a cavity floated in zero-$g$ can achieve the thermal-noise limit without sagging. This yields narrower laser linewidths for driving atomic transitions, and hence longer qubit $T_2$. One could even envision using the quiet space environment to run atoms on longer Raman or Ramsey sequences for quantum gates, taking advantage of the lack of vibrational interruptions. (Indeed, CAL showed that certain Ramsey interferometry schemes benefit from microgravity by allowing longer free-evolution times between laser pulses\cite{williams2024}.)

Empirical evidence strongly supports these advantages. During the CAL experiments, researchers observed that microgravity dramatically extended how long atoms could coherently evolve, enhancing the precision of quantum measurements and enabling experiments impossible on Earth\cite{roomISS2023}. For example, a Mach-Zehnder atom interferometer in orbit produced clear interference fringes, whereas a comparable apparatus on Earth of the same size would not have enough free-fall time for fringes to appear before the atoms hit the floor\cite{williams2024}. Similarly, drop-tower experiments (QUANTUS project in Germany) likewise demonstrated BEC formation and interference in $\sim$4.7 s of microgravity, refining cooling and interferometry techniques that directly leverage weightlessness\cite{williams2024}. In fact, the ability to adiabatically expand a BEC to extremely low kinetic energy was first explored in drop towers and later perfected on the ISS\cite{gaaloul2022}. Without gravity, atoms could be expanded to picokelvin energies, then recompressed or manipulated with minimal disturbance. These are precisely the conditions one would want for neutral-atom quantum processors or analog quantum simulators.

In summary, neutral atom qubits and quantum simulators likely see the most striking improvements in microgravity: coherence times can increase from milliseconds to seconds (or tens of seconds) as atoms float freely, negligible position drift allows the use of extremely weak traps and uniform arrays, and the ability to reach ultra-ultracold temperatures that stabilize qubits. Decoherence from mechanical noise is minimized since the entire experimental apparatus can be isolated in free-fall. The result is closer to the textbook ideal of a many-body quantum system evolving in near-perfect isolation. Indeed, as one review noted, space-based conditions improve both the precision of cold-atom quantum sensors and “the signal to be measured” by granting access to long interrogation times and cleaner environments\cite{bassi2022}. For quantum computing, the “signal” is the delicate phase and entanglement of qubits -- microgravity and cryogenics together protect that signal, allowing more operations and more complex algorithms before errors accumulate.

\section{Photonic Qubits in Space}
\textbf{Current Challenges:} Photonic qubits (quantum information encoded in photons, e.g. in their polarization, time-bin, or path) are somewhat different from matter-based qubits but are essential for quantum communication and certain computing architectures (such as linear optical quantum computing and measurement-based schemes). Photons do not suffer thermal decoherence in the same way atoms or electrons do; they travel at the speed of light and their main vulnerabilities are loss and phase noise. On Earth, a major challenge for photonic quantum information is loss during transmission (e.g. fiber attenuation or scattering in air) and environmental disturbances to optical phase (such as fiber thermal expansion or vibrational jitter in interferometers). Additionally, generating and detecting single photons with high efficiency often requires cryogenic devices (for example, superconducting nanowire single-photon detectors operate at $\sim$2~K, and certain nonlinear optical processes benefit from low temperatures to reduce noise). Conventional photonic quantum experiments are also limited by distance: sending entangled photons or qubits over more than on the order of 100~km of fiber is difficult due to absorption, and free-space terrestrial links are limited by line-of-sight and atmospheric turbulence.

\textbf{Space Benefits:} The most obvious advantage of going to space for photonic qubits is the ability to transmit quantum states over ultra-long distances without atmospheric loss. In orbit or deep space, photons can travel tens of thousands of kilometers in vacuum with only diffraction loss, enabling global-scale quantum communication links. This was spectacularly demonstrated by the Chinese \textit{Micius} satellite, which distributed entangled photon pairs between ground stations over 1200~km apart with sufficient fidelity to violate Bell's inequality\cite{yin2017}. Those experiments leveraged the orbital platform simply to get above the atmosphere; gravity itself did not directly affect the photons, but being in space removed a huge decoherence source (atmospheric scattering and turbulence). In essence, the space environment provides a pristine communication channel for photons that is impossible to achieve on the ground.

For photonic quantum computing hardware (such as optical circuits performing logic gates or generating cluster states), microgravity can offer subtler benefits. Consider an optical interferometer that is part of a photonic quantum gate: many linear optical gates require stable interferometers and phase references. On Earth, such an interferometer might drift due to vibrations or temperature gradients. In microgravity, one can achieve a much more stable interferometric setup. A spacecraft can be designed to have a very steady thermal environment (with multi-layer insulation and radiators) -- day/night cycles in low Earth orbit do cause temperature swings, but these can be mitigated. Additionally, without gravity, optical tables in a satellite can be mechanically isolated to an extreme degree (e.g. mounted on damping springs) because we do not have to counteract weight. This means alignment of optical components can remain rock solid over time. A laser beam can stay coupled into a fiber or waveguide with far less active feedback.

A recent milestone underscoring microgravity's advantages is the first quantum computer deployed in space: in 2025, an international team launched a small photonic quantum processor into low Earth orbit\cite{carpineti2025}. This device was built to survive rocket launch vibrations and then leverage the stable microgravity environment for operations. According to the team, the computer (a photonic chip) has been running in orbit, gathering data on how quantum circuits perform in space\cite{carpineti2025}. In space they cannot rely on constant human tuning or intervention, so the photonic system had to be passively stable. The successful deployment (the device survived launch and is operational) suggests that photonic circuits can indeed function in space, and once the violent launch is over, the operating environment is extremely calm. On Earth, photonic quantum computers require frequent calibration (fibers expand, components drift), but in orbit one cannot easily adjust them -- thus the stable conditions of microgravity actually facilitate long-term passive stability\cite{carpineti2025}. The team simplified their system and made it robust to thermal and mechanical shocks, essentially relying on the benign space environment for steady performance\cite{carpineti2025}. This illustrates a key microgravity advantage: one can have a long-lived, self-contained photonic system that does not suffer random jostling or gravitational sag of its optical components.

Another benefit of the space environment is electromagnetic cleanliness. A photonic quantum processor or communication node typically requires a low electromagnetic-noise environment (for instance, some single-photon detectors are sensitive to magnetic fields if they use superconducting transition-edge sensors or nanowires). In orbit, far from power mains and urban electromagnetic interference, the environment can be very quiet. Of course, a satellite has its own electronics, but these can be DC-powered or well-shielded. There is no pervasive 50/60~Hz power-line noise in a spacecraft unless intentionally introduced. Earth's ionosphere and Schumann resonances are also absent -- not that those strongly affect photonics, but it underlines the electromagnetic quietness of space.

Space-based photonic systems can also extend quantum memory times. Proposed orbital quantum repeaters, for example, might store photonic qubits in cold atomic gases or solid-state crystal memories; these could benefit from microgravity similarly to other atomic systems. If an orbiting quantum memory is free from gravity, the stored excitation (say, a spin wave in a cold atomic cloud) will not be disrupted by convection or by sedimentation of the medium. The memory device can also be cryogenically cooled by passive radiation to deep space, maintaining very low decoherence in the storage medium.

It should be noted that photons themselves are unaffected by a constant gravitational field in terms of their polarization or intrinsic coherence (ignoring the trivial gravitational redshift of their frequency). However, if photonic qubits take paths that traverse different gravitational potentials (e.g. one photon sent down to Earth and another stays on a satellite), general relativity predicts a relative phase shift between them. Experiments such as the planned Space QUEST mission aim to test whether this causes any decoherence or is simply a known phase shift\cite{bassi2022}. Standard quantum theory says it should not introduce randomness (just a calculable phase), but some alternative gravitational theories suggest gravity could induce decoherence in entangled photon pairs if the photons probe significantly different gravitational fields. Preliminary analyses indicate that any gravity-induced decoherence for entangled photons in low Earth orbit would be extremely weak\cite{bassi2022}. So far, results like those from \textit{Micius} show that entanglement can be preserved over orbital-scale distances, supporting the notion that space is an excellent medium for photonic qubits.

In summary, photonic qubits and optical quantum technologies benefit from microgravity largely through the absence of detrimental environment: no atmosphere, minimal vibration, stable thermal conditions, and the ability to cover planetary-scale distances. We do not speak of an increased “coherence time” for a photon (since a photon remains coherent until absorbed), but rather an improvement in interference visibility and in the fidelity of multi-photon operations. For example, a two-photon interference experiment will maintain higher visibility if the path lengths and phases are kept stable -- something microgravity helps ensure. The recent deployment of a photonic quantum computing experiment in orbit underscores that such systems can not only survive in space but perhaps outperform their terrestrial counterparts in stability\cite{carpineti2025}. One of the project leads, Philip Walther, remarked that he was proud the first quantum computer in space was developed to perform experiments in the extreme conditions of a space mission, pushing photonic technology to be versatile and robust\cite{carpineti2025}. This achievement suggests that future quantum networks might include nodes (quantum processors or repeaters) aboard satellites, benefiting from both ultracold infrastructure and microgravity to provide secure communication and distributed quantum computing.

\section{Evidence from Microgravity Experiments}
\textbf{Cold Atom Laboratory (ISS):} Deployed to the International Space Station in 2018, NASA's Cold Atom Lab (CAL) has produced quantum gases (BECs) in orbit and performed a variety of atom interferometry experiments. It demonstrated that microgravity allows effectively indefinite free-fall for atoms, leading to enhanced measurement precision and new phenomena observable that were impossible on the ground. For example, interference patterns from BECs persisted an order of magnitude longer than on Earth, confirming extended coherence times for matter waves in microgravity\cite{roomISS2023}. CAL also achieved record-low effective temperatures (on the order of 50--100~pK) via adiabatic expansion in free-fall\cite{gaaloul2022}. These capabilities translate directly to more coherent atomic qubits and sensors. (CAL investigators did note that vibrations on the ISS posed a challenge that needed mitigation\cite{swayne2024} -- implying that aside from such technical vibrations, the microgravity environment itself was beneficial once vibrational noise was controlled.) Upgrades to CAL (an improved Science Module 3 with a dedicated interferometer) have further pushed coherence times, enabling tests of fundamental physics (such as equivalence principle tests) with quantum gases\cite{alonso2022}. The takeaway from CAL is that microgravity works as hoped: experiments were done that are not possible on Earth, such as watching a freely expanding coherent cloud for several seconds\cite{roomISS2023}, thereby paving the way for space-based quantum technology.

\textbf{Drop Towers \& Parabolic Flights:} Prior to long-duration orbital experiments, drop towers (e.g. the 110~m ZARM tower in Bremen) and parabolic aircraft flights provided seconds-long microgravity windows for quantum tests. In the QUANTUS and MAIUS projects, rubidium BECs were created and studied during 4--9~s of microgravity\cite{williams2024}. Impressively, matter-wave interferometry was demonstrated in these conditions, and cooling protocols (like rapid evaporative cooling and delta-kick collimation) were refined to exploit the brief weightlessness\cite{williams2024}. The MAIUS-1 sounding rocket (2017) produced a BEC in space (over 6 minutes of microgravity) and conducted the first matter-wave interferometer in space\cite{williams2024}. These achievements proved that quantum coherence can survive launch and microgravity, and indeed benefit from it: the BECs in MAIUS showed no adverse effects from weightlessness; instead, they expanded symmetrically and remained trapped with minimal forces. The ability to perform 100+ drops per day in newer facilities (like the Einstein Elevator in Hannover) now allows iterative tests of quantum hardware in microgravity. For example, one can drop a vacuum chamber containing an ion trap or superconducting qubit and measure coherence over a 4~s free-fall, then compare results to static 1$g$ conditions. Though challenging, such experiments are on the horizon and will directly quantify microgravity's effect on qubit lifetimes. Already, groundwork indicates that neutral-atom coherence dramatically improves without gravity (in those drop experiments, BECs retained coherence longer than Earth-bound experiments of similar scale). Parabolic airplane flights have also been used to test quantum devices -- for instance, a recent parabolic flight carried a portable atom interferometer and found it worked reliably through repeated microgravity cycles. This suggests that quantum hardware can be robust and possibly gain performance in actual space missions.

\textbf{Space Clocks and Quantum Sensors:} As mentioned earlier, trapped-ion and cold atom clocks in space have matched or exceeded their Earth performance. For example, the $^{199}$Hg$^+$ ion Deep Space Atomic Clock (launched 2019) operated for over a year in orbit, maintaining a stability of roughly $3\times10^{-15}$ at one day\cite{alonso2022}. Its success demonstrated that microgravity did not introduce any new decoherence -- the clock's limiting noise was technical or due to reference oscillators, not gravity. Similarly, the cold rubidium clock on China's \textit{Tiangong-2} space lab achieved its stability goals in orbit\cite{alonso2022}, proving that laser-cooled atomic coherence (Ramsey fringe contrast, etc.) behaved as expected in free-fall. These are effectively single-qubit coherence demonstrations (clock stability reflects the coherence of a superposition over time). The results are encouraging: operating in space did not degrade coherence, and potentially even better performance could be attained if space advantages (no gravity, ultra-cold environment) are fully exploited. Another example is space-based quantum optical links: ESA's planned Space QUEST mission aims to test entangled photon transmission from the ISS to Earth to investigate gravitational effects on entanglement\cite{bassi2022}. Theoretical work suggests any gravity-induced decoherence will be extremely small\cite{bassi2022}. So far, experiments like \textit{Micius} have shown high-fidelity entanglement distribution, indicating that entangled photonic qubits remain coherent despite traveling through significantly different gravitational potentials (low Earth orbit vs ground)\cite{yin2017}. This is strong evidence that space conditions can preserve delicate quantum correlations over unprecedented scales.

\textbf{First Quantum Computer in Space (Photonic):} In June 2025, a small photonic quantum computing experiment (about the size of a shoebox) was launched into orbit\cite{carpineti2025}. As of this writing, detailed results are pending, but the mere fact that it was deployed and is operational is a milestone. The device is testing the behavior of photonic qubits (likely generating and manipulating entangled photons or running simple quantum algorithms) under space conditions. Engineers report that it had to be rugged and autonomous, but once in orbit it experiences a very stable thermal and mechanical environment\cite{carpineti2025}. If this experiment shows that gate fidelities or interference contrasts are as good as (or better than) on the ground, it will provide direct evidence supporting the hypothesis. Early statements from the team indicate they expect new innovations and applications to emerge from operating in such extreme conditions\cite{carpineti2025}. In other words, they suspect that some aspects of quantum operations might even improve, or that new techniques possible only in microgravity will be demonstrated.

All these examples reinforce the central idea: microgravity and ultracold conditions improve quantum coherence and reduce errors. So far, no fundamental obstacle has been observed in space -- quantum devices seem to be limited by the same factors as on Earth (materials, vacuum quality, control electronics), and when those are addressed, removing gravity only has positive effects on maintaining quantum states. The primary caution is that vibrations and other spacecraft-related noise must be carefully managed. The ISS, for instance, is not an ideal quiet lab (it vibrates due to crew activity, machinery, etc.), and CAL had to account for that\cite{swayne2024}. But that is a technical detail -- future free-flyer platforms can be made far quieter. In essence, the experiments to date show no show-stoppers and plenty of performance gains when leveraging microgravity.

\section{Proposed Experiment: Ground vs. Microgravity Quantum Processor Test}
To conclusively test the hypothesis that microgravity plus near-absolute-zero conditions enhance quantum computer performance, we propose a direct side-by-side experiment. The idea is to operate identical quantum computing hardware in two environments -- one on Earth (1$g$) and one in microgravity -- and compare key performance metrics: qubit coherence times ($T_1$ and $T_2$), gate fidelities, and readout error rates. By designing the setups to be as similar as possible in all other respects (temperature, shielding, control electronics), any differences in performance can be attributed largely to the presence or absence of gravity and associated factors.

\textbf{Hardware and Platforms:} We choose a platform that is compact and mature enough to deploy: for example, a trapped-ion quantum processor with 2--4 qubits, or a superconducting qubit module with a few qubits and readout resonators. Trapped ions are an attractive choice because they have very long coherence and are sensitive to environmental noise (so improvements would be measurable), yet a small ion trap system can be made transportable. Superconducting qubits would require a dilution refrigerator, which is heavier, but emerging compact dilution fridges and closed-cycle cryocoolers could make space deployment feasible. For concreteness, suppose we use a trapped-ion system (since space agencies have experience flying atomic clocks, which use similar hardware). The system would consist of a miniaturized ion trap (with, say, $^{171}$Yb$^+$ ions), lasers or microwave sources for gating, and a vacuum/cryogenic housing.

The ground unit would be set up in a laboratory with the ion trap operating at 4~K (to mirror the space unit's thermal environment). It would be enclosed in magnetic shielding and placed on a vibration-isolated optical table to create the best possible environment on Earth. The space unit would essentially be the same ion trap in a 4~K cryostat, with the same laser/control system, mounted either on a small satellite or as an experiment module on the ISS. Crucially, after launch the space unit will be in microgravity and can be further isolated (for example, mounted on a passive vibration damping stage). Both units would use identical control sequences and software to run quantum logic gates and benchmarking routines (such as randomized benchmarking for gate fidelity, and Ramsey/spin-echo experiments for coherence times).

\textbf{Experimental Procedure:} On both systems, we would perform a series of tests: (1) Measure individual qubit coherence ($T_1$ relaxation and $T_2$ Ramsey decoherence) under identical pulse sequences. (2) Execute single-qubit rotations and two-qubit entangling gates, using quantum process tomography or randomized benchmarking to evaluate their fidelities. (3) Prepare a Bell state (two-qubit maximally entangled state) and measure its longevity (how long entanglement is preserved) in each environment. (4) Perform repeated rounds of qubit readout to assess measurement accuracy and look for any drift in qubit frequencies that might indicate environmental perturbations.

These tests would run continuously or periodically. Results from the space system would be telemetered to Earth for comparison. The space unit could be on the ISS (for convenient power and data links) or on a free-flying small satellite. The ISS environment offers microgravity but has some vibration; a free-flyer could be equipped with an active inertial platform to counteract any internal vibrations and achieve very smooth free-fall. If using the ISS, one might schedule runs during quiet periods to minimize vibration. The ground unit, meanwhile, would be in a stable lab but of course subject to 1$g$ and unavoidable micro-seismic vibrations.

\textbf{Isolating Gravitational Effects:} The biggest challenge is ensuring that any observed differences are due to microgravity and not other disparities between the setups. We address this by strict configuration matching and monitoring of extraneous factors. Both systems would carry environmental sensors: accelerometers to measure vibrations, magnetometers for stray fields, and particle detectors for radiation events. This way, if the space unit shows longer coherence, we can confirm it was not simply because the space unit was quieter in terms of vibration or magnetic noise (although microgravity indirectly allows a quieter setting, we could also attempt to match the ground unit's vibration spectrum to the space unit's if possible). In practice, the Earth lab will always have 1$g$, but it can be extremely vibration-isolated (perhaps even placed underground to reduce seismic noise). The space unit will have 0$g$ but some vibrations from the spacecraft. By monitoring acceleration, we can later weight or filter the results to account for specific vibration events. Similarly, both units would operate at the same temperature (4~K or whatever is chosen) to ensure thermal noise is equal. The space unit might experience more cosmic ray hits; we can partially mitigate that by adding radiation shielding on the satellite until the cosmic ray flux is similar to ground level (though perfect equivalence is hard). Alternatively, we include a detector to count cosmic ray events on the qubits (e.g., a coincident error signal or a dedicated sensor) so that if the space unit suffers more such events, those data points can be excluded for fair comparison. We could also choose a high orbit with minimal charged-particle flux.

Another approach to isolate gravity's role is to perform a differential measurement by modulating the effective gravity for the space qubits and observing any change in coherence. This could be done by intentionally accelerating the spacecraft slightly (introducing a small pseudo-gravity) or by using a centrifuge on board to provide a small $g$-force on a duplicate set of qubits. For example, one could spin up part of the apparatus to simulate, say, 0.01~$g$ for a control run in space, then spin it down to 0~$g$. This would directly toggle the presence of gravity. The expectation is that even a small effective $g$ might start to introduce more decoherence, strengthening the evidence that gravity (or its absence) is the key factor. Such a gravity-modulation experiment is complex but conceivable on a small satellite with a rotation stage.

\textbf{Expected Outcomes:} Based on the earlier reasoning, we anticipate the space (microgravity) quantum processor to show measurable improvements. For instance, the Ramsey $T_2$ of an ion qubit might be longer in space -- perhaps limited only by magnetic noise and the ion's intrinsic properties, instead of by vibrations or blackbody shifts. If on Earth $T_2$ is, say, 1~s with best efforts, in space we might see it extend to several seconds, indicating fewer environmental phase perturbations. Gate fidelities on Earth might plateau due to beam pointing stability limits or mechanical vibrations (e.g. stuck around 99.9\%), whereas in space they might approach the fundamental limits set by quantum noise (perhaps 99.99\% or better). An entangled Bell state might lose 50\% of its fidelity after 500~ms on Earth (due to accumulated dephasing), whereas in microgravity it could maintain high fidelity for multiple seconds. Readout error rates might also improve if the detection optics are more stable (in space there is no 50~Hz mains electrical noise, and the photodetectors might see lower background counts due to the cosmic dark sky -- although one must consider cosmic ray hits causing spurious counts).

Statistical analysis of the data would determine if improvements are significant. For example, if the microgravity qubit achieves a $T_2$ twice as long as the lab qubit under identical conditions, that would be strong evidence that the absence of gravity (and its associated effects) improved coherence. If no improvement is seen, it might imply we had already eliminated gravity-related noise on the ground or that other factors dominate (meaning microgravity alone is not a panacea unless accompanied by better shielding, etc.). However, given prior experiments, we expect to see differences. Notably, we could include deliberately gravity-sensitive metrics -- for instance, the gravitational redshift phase drift. By putting two ions at slightly different heights in the trap and measuring their relative phase evolution (comparing 1$g$ on Earth vs 0$g$ in space), we could directly confirm the predicted gravitational dephasing effect\cite{balatsky2025}. On Earth, over long times a tiny phase drift would accumulate between them, whereas in space it would not. Such a measurement would explicitly verify gravity's contribution to dephasing as predicted by theory.

\textbf{Potential Issues:} We must acknowledge various challenges. Launching delicate quantum hardware means exposure to intense vibrations and shocks (which can be mitigated by robust design and packaging -- as demonstrated by the photonic experiment's successful launch\cite{carpineti2025}). Maintaining cryogenics in space is non-trivial but feasible using closed-cycle cryocoolers (note that the space unit's cryocooler could introduce microphonic vibrations -- ironically reintroducing some noise; one solution is to use magnetically driven or sorption-based coolers with fewer moving parts). Additionally, cosmic radiation might cause more qubit “blips” in space; if that noise dominates, it could mask the improvements in baseline coherence. Careful shielding and/or event veto schemes (as mentioned) would be needed to discount radiation-induced errors.

\textbf{Gravity Modulation Tests:} To isolate gravity itself (as opposed to just being "in space"), we consider performing the experiment under different gravitational conditions: (a) Spacecraft in free-fall (0$g$), (b) Spacecraft accelerating (simulating a partial $g$), and (c) Ground (1$g$). If the main performance difference is between the 1$g$ and 0$g$ cases, that strongly indicates gravity (or its absence) made the difference. If an intermediate partial-$g$ case shows intermediate performance, that's a smoking gun for gravitational scaling of decoherence. We also ensure both locations have similar magnetic fields (we can zero the local field with coils in both setups) so that, for instance, Earth's 50~µT field vs. an orbital environment (maybe 30~µT at ISS altitude) doesn't confound the results -- both traps can be actively zeroed to micro-tesla levels internally. Both systems use identical microwave and laser controls, so technical noise is comparable.

In short, this proposed experiment is akin to a “quantum computing twin test” (analogous to NASA's twin astronaut experiment in which one twin went to space and one stayed on Earth). Here we would send one quantum computer to space and keep an identical one on Earth. By analyzing their performance divergence, we can directly assess the claim that microgravity plus ultracold isolation yields better quantum computing conditions. This would provide the ultimate validation of all the individual pieces of evidence described earlier.

\section{Conclusion}
Extreme environments breed extreme performance -- this appears to hold true for quantum computing hardware just as it does in other physical domains. The convergence of microgravity and near-absolute-zero temperature offers a glimpse of the idealized frictionless world that quantum engineers strive for: qubits evolving undisturbed by stray forces, perturbations, or thermal randomness. Through theoretical arguments and an accumulating body of experimental demonstrations, we have argued that removing the shackles of gravity and thermal noise does more than just solve the obvious problem of things “falling down.” It fundamentally suppresses multiple decoherence channels -- from gravitational phase shifts to vibration-induced noise to convective disturbances -- thereby improving coherence times and operational fidelities across superconducting, trapped-ion, neutral atom, and photonic quantum platforms.

In superconducting circuits, weightlessness and a cryogenic vacuum promise quieter electromagnetic environments and freedom from microphonic resonator shifts, potentially allowing qubit coherence to reach the intrinsic material limits. In trapped-ion systems, microgravity permits weaker, more pristine trapping and eliminates the need to fight gravity, yielding longer ion-string coherence and higher gate fidelity in a gentler, ultra-high vacuum setting. Ultracold neutral atom architectures arguably benefit the most: microgravity frees them from the millisecond timescales of free-fall, extending coherence to seconds and allowing qubits to be manipulated in perfectly symmetric traps at picokelvin temperatures -- conditions unattainable on Earth. Photonic qubits find in space a lossless, stable propagation medium and a serene platform for interference experiments, paving the way for global quantum networks and perhaps even orbiting quantum processors. Across the board, the trend is clear: the harsher (or more unusual) the environment for humans, the more hospitable it becomes for delicate quantum information.

The performance boosts discussed here are not merely speculative. Space-based quantum clocks have shown that long-lived superpositions can survive and even excel in orbit. Bose--Einstein condensates and matter-wave interferometers in microgravity have maintained coherence far beyond terrestrial limits, providing direct evidence that quantum states truly flourish when gravity is turned off\cite{roomISS2023,williams2024}. The launch of the first quantum computing experiment to space in 2025\cite{carpineti2025} marks the start of a new era -- one in which we will empirically validate how quantum logic behaves in extraterrestrial settings. If the hypothesis holds, we may soon see quantum computers deployed as payloads on space stations, free-flyers, or lunar bases, not just for novelty, but because those environments genuinely optimize their performance.

Of course, realizing this vision will require overcoming engineering hurdles. Space is a challenging operating theater: radiation, vacuum, and remoteness demand robust, autonomous systems. Yet the very process of adapting quantum hardware for space yields designs that are more compact, rugged, and less error-prone -- qualities that also benefit terrestrial systems. There is a symbiosis: pursuing space-based quantum computing will drive improvements in miniaturization and error correction that feed back to Earth, while the unique features of space (like microgravity) feed forward to enhance capabilities. Looking ahead, one can imagine hybrid quantum networks where ground and orbital qubit processors work in tandem, linked by photonic qubits beaming through space, with the orbital nodes enjoying superior coherence as “quantum hubs” hovering above the noisy Earth.

In conclusion, microgravity and ultracold conditions do more than solve gravitational inconveniences; they create a closer approximation to the ideal isolated quantum system, thereby improving the performance of quantum computing hardware at a deep level. The hypothesis that these environments simulate ideal computational conditions is strongly supported by theoretical models (treating gravity as a decohering noise and basic thermodynamic arguments) and is increasingly corroborated by experimental milestones from drop towers to the ISS\cite{gaaloul2022,williams2024}. As our proposed comparative experiment suggests, the ultimate proof will come from head-to-head tests -- and if those confirm the advantages, it could herald a paradigm shift: the highest-performance quantum computers might eventually reside in orbit or deep space, where Mother Nature provides the refrigeration and silence that quantum coherence craves. Far from Earth's pull and at the edge of absolute zero, we may find the true potential of quantum computation unlocked, tackling classically intractable problems under the quiet stare of the stars.

\section*{Acknowledgments}
The author thanks colleagues at Robotech Frontier Hub for insightful discussions. This work was supported by the Robotech Frontier Hub internal research program.

\bibliographystyle{unsrt}
\bibliography{microgravity_quantum}
\end{document}